\documentclass[sigconf]{acmart}

\definecolor{brown}{rgb}{0.59, 0.29, 0.0}
\definecolor{darkgray}{rgb}{0.59, 0.59, 0.59}
\definecolor{tablegray}{gray}{.9}

\newcommand\rv[1]{\textcolor{black}{#1}}
\newcommand\remove[1]{}

\usepackage[utf8]{inputenc}
\usepackage{colortbl}

\usepackage{color}
\usepackage{booktabs}
\usepackage{multirow}

\usepackage{xspace}
\usepackage{enumitem}
\usepackage{algorithm}

\usepackage{amssymb}
\usepackage{pifont}
\usepackage{amsmath}

\newcommand{\customtilde}{{\raise.17ex\hbox{$\scriptstyle\sim$}}}

\newcommand{\etal}{et~al.\xspace}
\newcommand{\eg}{e.\,g.,\xspace}
\newcommand{\ie}{i.\,e.,\xspace}

\AtBeginDocument{%
  }

\copyrightyear{2026}
\acmYear{2026}
\setcopyright{cc}
\setcctype{by}
\acmConference[UIST '26]{The 39th Annual ACM Symposium on User Interface Software and Technology}{November 02--05, 2026}{Detroit, MI, USA}
\acmBooktitle{The 39th Annual ACM Symposium on User Interface Software and Technology (UIST '26), November 02--05, 2026, Detroit, MI, USA}
\acmDOI{10.1145/3830398.3830530}
\acmISBN{979-8-4007-2856-3/2026/11}

\usepackage{xspace}
\usepackage{xcolor}

\newcommand{\projectname}{MultiVerse\xspace}
\newcommand{\baseline}{Prompt\xspace}
\newcommand{\approach}{$C^3$\xspace}
\newcommand{\persona}{Anna\xspace}

\usepackage{booktabs}

\begin{document}

\title{MultiVerse: A Creator-Centered Approach to Steering Context-Adaptive Lyrics}


\author{Alexander Wang}
\affiliation{%
  \institution{\mbox{Human-Computer Interaction Institute}}
  \institution{Carnegie Mellon University}
  \city{Pittsburgh}
  \state{Pennsylvania}
  \country{USA}}

\author{Chris Donahue}
\affiliation{%
  \institution{Computer Science Department}
  \institution{Carnegie Mellon University}
  \city{Pittsburgh}
  \state{Pennsylvania}
  \country{USA}}

\author{David Lindlbauer}
\affiliation{%
  \institution{\mbox{Human-Computer Interaction Institute}}
  \institution{Carnegie Mellon University}
  \city{Pittsburgh}
  \state{Pennsylvania}
  \country{USA}}

\renewcommand{\shortauthors}{Wang et al.}

\begin{abstract}
Generative AI may enable new forms of context-aware creative expression by dynamically tailoring media content to its consumption context.
For instance, AI systems could adapt song lyrics to the listener and their current activity.
However, existing media adaptation systems primarily optimize for audience experience, often neglecting artists' intent, style, and preference.
We address this challenge by introducing a novel creator-centered approach to adaptive media authoring and present \projectname, a system that instantiates this approach for steering adaptive lyrics.
Our approach allows creators to explicitly author controls based on their intent, lyric structure, and audience context, and uses rule-based validations to ensure controls are followed.
We conducted a study with 10 songwriters, comparing \projectname with a prompting-based workflow for composing adaptive lyrics. 
\remove{Songwriters associated \projectname with greater control and predictability in lyric adaptation, and valued its audience-centric features for producing genuinely personalized lyrics.
Our approach balances adaptability and control by focusing on creator-facing tools for authoring context-adaptive media.}
\rv{The comparison revealed that creators preferred to author how lyrics adapt by explicitly specifying relevant context and adaptation constraints, while recognizing trade-offs in flexibility and iteration speed. Interviews further revealed that creators viewed adaptive media as enabling new forms of audience connection, introducing a distinct creative process, favoring new compositional strategies, and reshaping notions of authorship.}

\end{abstract}





\begin{teaserfigure}
  \includegraphics[width=\textwidth]{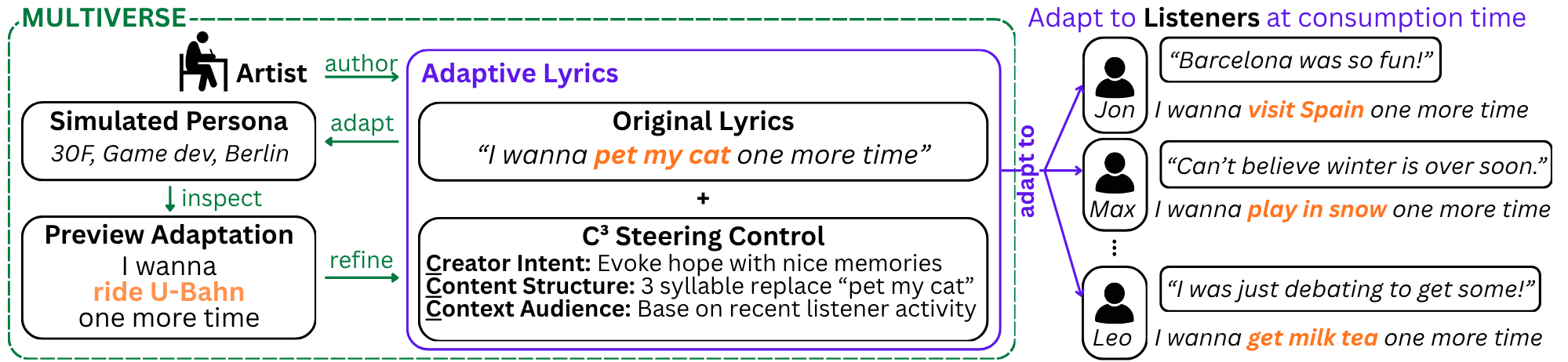}
  \caption{
  \rv{We anticipate the rise of AI-driven trends in music and multimedia like \emph{adaptive lyrics}, where lyrics are dynamically tailored to} different audiences at consumption time. Because final outputs are generated at consumption time, artists cannot directly review or edit each variation, \rv{running the risk of deviating from their creative intent}. MultiVerse addresses this by providing tools that allow artists to iteratively author steering controls over intent, content structure, and audience context, while previewing how lyrics may adapt across audiences with simulated personas. Once the controls are refined, they can be deployed to deliver adaptive lyrics to real listeners at scale.
  }
  \label{fig:1}
\end{teaserfigure}


\maketitle

\section{Introduction}

Consider the following example: A listener is boarding a late-night flight to start a new job abroad, leaving behind their hometown and family. As they sit at the gate listening to\textit{ Don’t Stop Believin’} by Journey, the original verse (“Just a city boy, born and raised in South Detroit. He took the midnight train goin’ anywhere”) \textit{adapts} to reflect their situation:
“Just a quiet girl, leaving home in search of more. She took a red-eye flight bound for San Jose.”
The modified lyric mirrors the listener’s circumstances, making the song feel more personally relevant—as if the song briefly breaks the fourth wall, allowing the artist to directly address the listener’s own moment of transition.

Adaptive lyrics, like in the earlier example, demonstrate how songs can respond to individual listeners to create a more personal experience. In practice, artists have long adapted their work to engage different audiences.
At a broad scale, they create region-specific editions with translated or culturally localized content (\eg Nena's \textit{99 Luftballons} and \textit{99 Red Balloons}), and clean radio edits that tone down profanity and violence. At a finer scale, they often tailor certain lines during live shows to reference shared experiences with the audience. However, this quickly becomes impractical to scale across the countless contexts in which listeners encounter music.
 
Recent advances in AI, however, are breaking through that barrier—enabling music to be automatically adapted in real-time based on the listening context.
Early work in this space has primarily focused on listener-side benefits, such as reducing disruption~\cite{wang2024maringba}, improving awareness~\cite{kari2021soundsride}, and boosting exercise motivation~\cite{wang2025rise}.
Notably, Wang~\etal introduced a method that modifies song lyrics to reflect real-time notifications~\cite{wang2024towards}—\eg replacing original lyrics with “your flight to NY is boarding now.”
Adaptive lyrics could thus adjust automatically to individual audience members, \eg their preference, location, or aspects of their personality.
This personalization can increase the connection between artists and their audience, and improve the experience for the audience~\cite{cla}.
This approach, however, comes at the cost of taking control from artists, and potentially deviating from their intent.
It is thus unclear, how artists can retain control over AI adaptations of their original work. 

\rv{Motivated by prior formative work on creator perspectives \cite{cla}, }we highlight two main challenges of adaptive media.
First, artists need ways to \textit{control} how their creations are adapted to audience members. 
This includes specifying which parts could be modified and which parts must remain untouched, or controlling how aspects of an audience member should be infused into the medium.
Second, since adaptive lyrics are generated at consumption time, artists cannot fully anticipate how each audience member will experience their work. 
Artists thus need ways to systematically preview how their original work is adapted for audience members, and take this into account for steering the adaptive media.

To address this, we propose \approach, a new creator-centered paradigm for adaptive media authoring, where artists explicitly steer how their work responds to consumption context rather than fully delegating adaptation to automated systems (Figure~\ref{fig:1}). 
Our approach is grounded in three design axioms corresponding to three factors in adaptive media: creator, content, and context. Rather than handling these factors implicitly within automated generation pipelines, our approach exposes them as explicit dimensions for steering. Creators specify how their work aligns with their intent (creator), define structural constraints that must be preserved across variations (content), and articulate which consumption-side conditions are relevant and how adaptation unfolds in response to them (context).

To operationalize \approach and examine its implications in practice, we instantiated this approach in the domain of songwriting through the development of a tool for steering adaptive lyric content. We contribute \projectname, an adaptive lyric steering system that exposes creator, content, and context as structured controls for lyric adaptation. 
\projectname explicitly exposes context as simulated personas for authors to preview how lyrics adapt to different audiences, how they react to the lyrics, and allow creators to proactively query for specific audience information to shape adaptation behavior. 
\projectname implements domain-specific tools to control content structure such as annotating the desired stress pattern, syllable count, and which words should rhyme. 
The system further incorporates rule-based validation mechanisms that check generated lyrics against musical and structural requirements, ensuring coherence while constraining the space of possible variations.

\vspace{-1em}
\paragraph{Envisioned usage}
We envision that artists leverage \projectname to adapt their \textit{existing lyrics}, as outlined in the walkthrough (Sec.~\ref{sec:walkthrough}). Given original lyrics, they author \textit{steering controls} that guide how lyrics can be adapted for different audience members. MultiVerse allows artists to preview adaptations for simulated personas and see how those personas might respond to the lyrics. 
The \textit{output} of \projectname is a set of \textit{steering controls}, defined for use in conjunction with the original lyrics.
When audience members listen to the songs, a future system (e.g., a plugin for Spotify or Apple Music) would adjust the lyrics based on these \textit{steering controls}. Contextual information could be provided directly by listeners (e.g., demographics, occupation, favorite food) or extracted automatically by other systems (e.g., location). This would allow listeners to experience personalized versions of songs, such as hearing their favorite artists sing about the city they live in.
In this paper, we focus on exploring how artists maintain control over their work, leaving the study of audience perceptions and preferences for future work.

To understand how \approach and \projectname intersect with creative practice, we contribute insights from a study with 10 songwriters. 
Participants transformed their own compositions into adaptive versions using two contrasting workflows: \projectname, and a prompt-based LLM workflow representative of contemporary generative lyric practices. 

\rv{Our study provides two complementary sets of insights: first, comparing MultiVerse with prompting revealed that creators valued mechanisms for defining meaningful adaptation context and explicit control over adaptations, while recognizing trade-offs in adaptation flexibility and iteration speed; second, interviews revealed broader creator perspectives on adaptive media, showing how it enables new forms of audience connection, introduces distinct creative processes and compositional strategies, and reshapes notions of collaboration, authorship, and ownership.}

In summary, our work contributes: 
\begin{itemize}[leftmargin=*, align=left, label={}]
    \item \textbf{(1) $\boldsymbol{C^3}$, a creator-centered authoring
    approach for adaptive media} that exposes \underline{c}reator, \underline{c}ontent, and \underline{c}ontext as explicit dimensions in the authoring workflow.

    \item \textbf{(2) MultiVerse, 
    an adaptive lyric steering system} that operationalizes \approach through structured controls and rule-based validation of adaptations.

    \item \textbf{(3) A study with songwriters (N = 10)}, providing empirical insights into how creators engage with \approach and the trade-offs between explicit structured control and prompt-based workflows.
\end{itemize}

\section{Walkthrough}
\label{sec:walkthrough}
To better illustrate \approach and its instantiation in \projectname for adaptive lyrics, we provide a walkthrough of a creator iteratively authoring steering controls. Suppose a creator is using \projectname to author steering controls. Their original lyrics are "\textit{"My dog just died my friend hates me, I saw myself on MTV"}"  and their goal is to have them adapt dynamically to each listener’s context — while respecting a few key constraints: (i) maintaining the original rhyme scheme, (ii) preserving the theme of isolation and feeling lost, and (iii) connecting to audience contexts that resonate with that theme. They begin by focusing on a single example user persona, then validate the pipeline across a broader set of audiences.

\begin{figure}[H]
  \centering
  \caption{Adapt for persona \persona with no constraints.}
  \includegraphics[width=.9\linewidth]
  {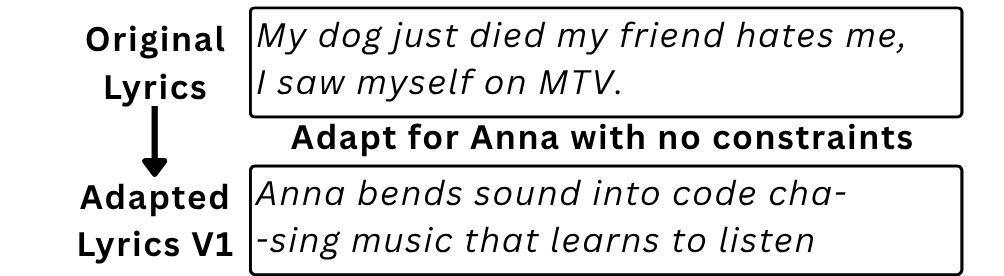}  
  \label{fig:a}
\end{figure}

\textbf{No constraint (Figure \ref{fig:a}).}
The creator starts by adapting lyrics with no controls
for \textit{\persona}, a young graduate student studying music and CS, they observe the lyric: \textit{"\persona bends sound into code, chasing music that learns to listen,"} which shifts focus entirely to the persona, producing a line that no longer reflects the artist’s original intent.

\vspace{-2em}

\begin{figure}[H]
  \centering
  \caption{Artist specifies content structure.}
  \includegraphics[width=.9\linewidth]{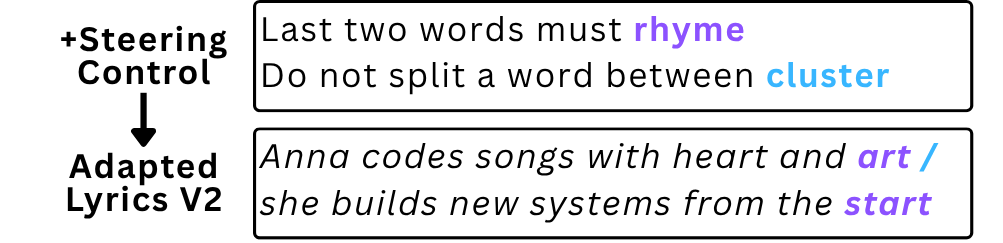}
  
  \label{fig:c}
\end{figure}

\textbf{Content structure (Figure \ref{fig:c}). }
In the original melody, there is a long pause after the word "me" (1 syllable), which is currently being replaced with "chasing", a 2 syllable word that now splits across two clusters of rhythm. Furthermore, "me" and "V" in the original lyrics rhyme. To fix this, the creator uses the rhyme tool to bind note 8 with note 16, and the cluster tool to split note 8 with 9. 

Using the updated control settings, the new output becomes: \textit{"\persona codes songs with heart and art, she builds new systems from the start."} The lyrics now fit the original composition, but still diverge from the theme of the song.

\vspace{-1em}

\begin{figure}[H]
  \centering
  \caption{Artist specifies creator intent.
}
  \includegraphics[width=.9\linewidth]{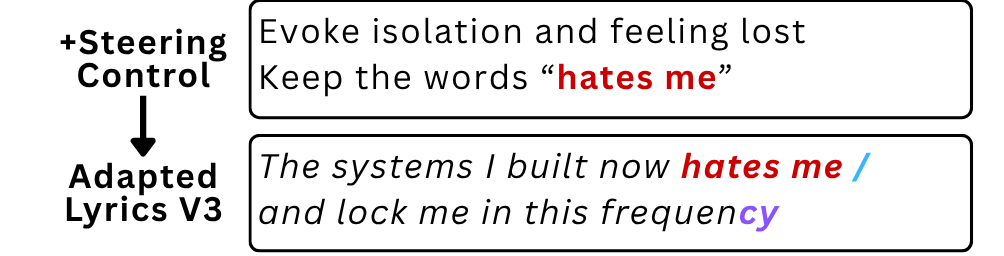}
    \label{fig:d}
\end{figure}

\textbf{Creator intent (Figure \ref{fig:d}). }
When initially composing the song, the creator intended the music to convey themes of isolation and feeling lost, and this specific line should convey the feeling of being rejected by those around them. To fix this, the creator uses global intent to specify the song's theme and the word lock to ensure that the output keeps the exact words "hates me." 

Using the updated control settings, the new output becomes: \textit{"The system I built now hates me, and lock me in this frequency."} The lyrics now fit the overall theme of the song, but still feels artificial and only adapt based on superficial aspects of the listener.

\vspace{-1em}

\begin{figure}[H]
  \centering
  \caption{Artist specifies relevant audience context.
}
  \includegraphics[width=.9\linewidth]{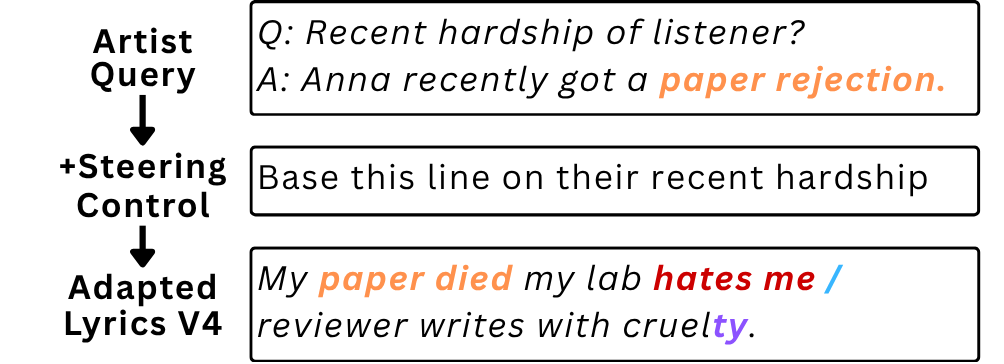}
  
  \label{fig:e}
\end{figure}

\textbf{Audience context (Figure \ref{fig:e}). }
To effectively tailor the content to \persona, the virtual persona, the creator uses the contextual query tool to proactively ask specific questions about the listener and reveal key contexts relevant to the music. The creator asks, "what happened recently that makes you feel lost?", and \projectname responds that "A recent hardship for \persona is having their paper rejected. The harsh comments from reviewers and isolation from their research lab leave them feeling lost."

With the updated audience context, the new output becomes:\textit{ "My paper died my lab hates me, reviewer writes with cruelty."} 
The creator is now satisfied with this output.

\vspace{-0.5em}
\begin{figure}[H]
  \centering
  \caption{Artist preview output for other personas.
}
  \includegraphics[width=.9\linewidth]{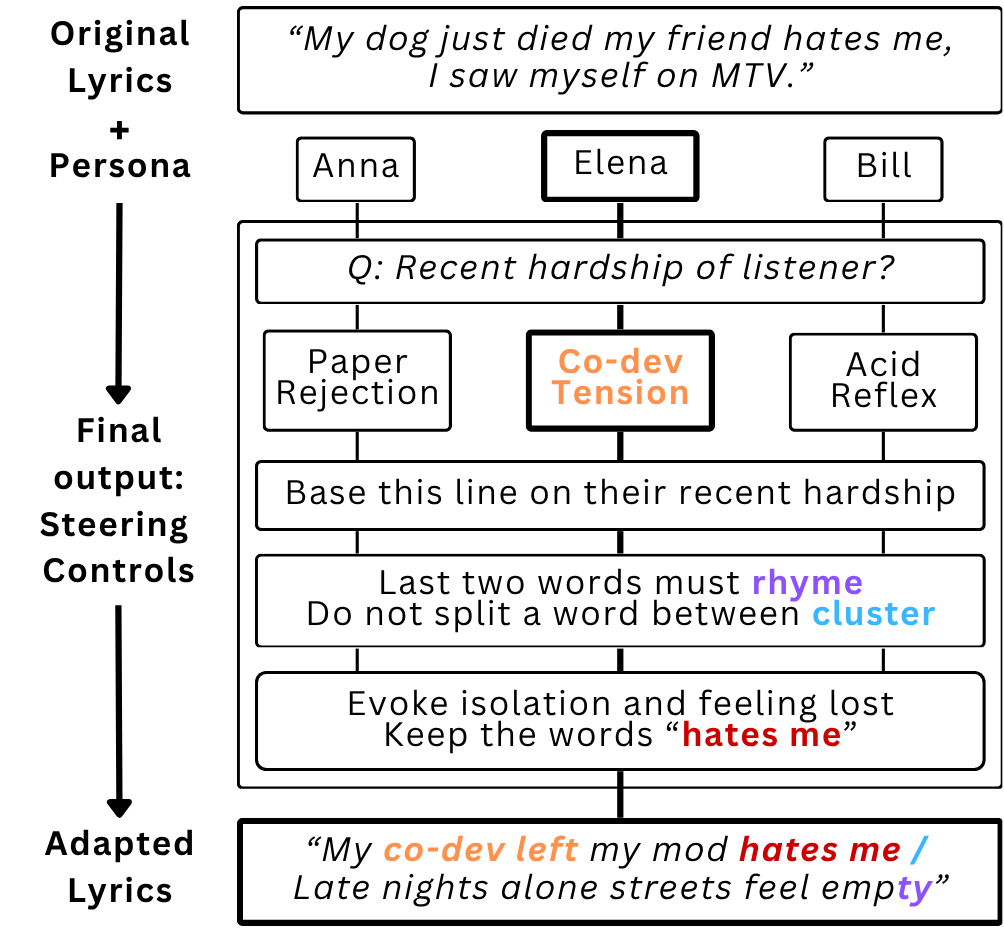}
  
  \label{fig:f}
\end{figure}

\vspace{-0.5em}
\textbf{Generalizing across audiences (Figure \ref{fig:f}). }
While the lyrics for \persona are satisfactory, the creator next evaluates how well the same set of steering controls generalize across several virtual personas. For example, Elena is an indie game designer in Berlin who collaborates closely with online creative communities and helps run a Discord server for her game. Through contextual query, the system simulates how the authored controls would adapt the lyrics given her personal context—such as tension with a collaborator—while preserving structural and thematic constraints. The creator repeats this process for multiple personas, making minor adjustments whenever unexpected adaptations arise. 
They confirm that the steering controls produce adaptations that are well aligned with their intent, and finalize them for application to the original lyrics, enabling the song to dynamically adapt for real listeners at scale.

\vspace{-2em}
\section{Related Work}

\subsection{Generative AI for media}

Generative AI is now capable of producing multimedia with high fidelity and controllability, and increasingly allows fine-grained edits to existing media content, enabling systems to dynamically adapt media during consumption. Notably, commercial tools such as Google Gemini (Nano Banana~\cite{nnbnn}) allow users to freely manipulate images via text prompts, enabling operations such as adding or removing objects, altering character poses, and transforming visual style with fine-grained control.

Similarly, recent models in the music domain also demonstrate notable advances in fine-grained control ~\cite{wu2024music, wu2023large}, real-time inference ~\cite{caillon2021rave, engelddsp}, content infilling ~\cite{thickstunanticipatory, huang2019counterpoint}, and editing ~\cite{tsai2024audio}. For instance, ACE-step enables modifying input music to change the singer, style, and lyrics ~\cite{gong2025ace}. While these capabilities are still emerging, they signal a broader shift: the kind of open-ended manipulation already possible in images is soon coming to music and other creative domains. This rapid rise of malleable content creates a pressing need for frameworks that give creators control over media adaptation.

\subsection{Context-aware interfaces and media}

Computing systems are increasingly able to sense and respond to the context in which they operate ~\cite{dey2001understanding}, enabling digital experiences that adapt to the surrounding physical environment and user state and actions ~\cite{jones2013illumiroom, kari2023scene, rajaram2024blendscape}. 
While substantial prior work focuses on adapting interface elements without altering the underlying content~\cite{han2023blendmr, cheng2021semanticadapt, cho2024auptimize, evangelista2022auit}, context-awareness is applied to modify the content itself, shaping storytelling media across modalities such as VR~\cite{tao2022integrating}, film~\cite{heck2021subconscious}, video captions~\cite{huang2025captune}, books~\cite{gunturu2024augmented},~\rv{ journalism ~\cite{10.1145/3025453.3025631}, and scientific communication ~\cite{10.1145/3589955,10.1145/3757660}.}

Similar context-aware adaptations have also been explored in music. Early work primarily focused on recommending songs suited to contextual factors such as mood, location, cultural background\rv{, and conversations} ~\cite{10.1145/2502081.2502170, schedl2014location, zangerle2020user, 10.1145/3209219.3209258, 10.1145/3359181}. 
More recent work adapts the music itself, including structure-level modifications that align musical transitions with activities such as driving or weightlifting ~\cite{kari2021soundsride, wang2025rise}, as well as content-level changes that alter existing song elements, such as integrating ringtones as new melodies or replacing lyrics with speech from conversational agents~\cite{wang2024maringba, wang2024towards}.
Currently, prior work in context-adaptive systems either adapts interface elements without considering the underlying content, or modifies content automatically without regarding creator’s intent (Figure \ref{fig:venn}).

\subsection{Tools for authoring static media}
Generative AI has increasingly been explored as a creativity support tool for producing many forms of static media content, including text ~\cite{writingassistant, gero2022sparks, 10.1145/3715336.3735832, 10.1145/3772318.3790568}, image ~\cite{lin2025inkspire, brade2023promptify, 10.1145/3563657.3595961}, and video ~\cite{wang2024lave, huh2025videodiff, wang2026rewriting, liu2026text, 10.1145/3379337.3415845}. In music, AI-assisted systems support artists in iteratively exploring ideas, refining outputs, and collaborating with generative models, accelerating ideation while raising questions around creative agency and authorship ~\cite{newman2023human, 10.1145/3715336.3735829}. Prior work introduces mechanisms for steering generation ~\cite{louie2020novice, louie2022expressive, 10.1145/3706598.3713861}, incorporating multimodal inspirations and queries ~\cite{10.1145/3706598.3713818, brade2024synthscribe}, and expressing abstract moods to guide outputs ~\cite{10.1145/3715336.3735814}. Generative approaches have also been applied to songwriting, including generating lyrics under structural constraints and composing melodies aligned with existing lyrics ~\cite{10.1145/3025171.3025194, watanabe-etal-2018-melody, sun-etal-2023-songrewriter, donahue-etal-2020-enabling, lee-etal-2019-icomposer, wang2024towards, 10.1007/978-3-319-55750-2_1}.

While prior work focuses on improving creator control over static generated outputs, our setting departs from this assumption: the final output is not fixed, but dynamically generated at consumption time. As a result, creators cannot directly inspect or anticipate all possible variations, necessitating new approaches to encoding and preserving artistic intent.

\subsection{Tools for authoring dynamic media}

Currently, the most mature ecosystem for creator-authored dynamic media exists in interactive content, such as video games and interactive narrative systems that allow players to progress through stories non-linearly. These systems typically adopt branching narrative structures, in which player choices route them along pre-authored paths toward distinct narrative outcomes ~\cite{1626183}. Authoring tools for such systems often rely on node-based environments (e.g., Twine, Ink, ChoiceScript), enabling creators to script conditional logic and manage narrative branches. 

Prior work has further explored systems that generate narrative content and gameplay behavior at runtime ~\cite{10.1145/3337722.3337732, 10.1145/3402942.3409599, 10.1145/3654777.3676358}, moving beyond fixed branching structures. These approaches demonstrate how generative models can alter storylines dynamically based on player interaction, enabling personalization that is no longer constrained to a small set of predefined paths. This shift enables a form of authoring centered on world planning ~\cite{10.1145/3654777.3676352, lebowitz1985story}, defining individual agent behavior ~\cite{park2023generative, 10.1145/3649921.3656987}, and constraining the space of possible narrative trajectories ~\cite{riedl2006story, 10.1145/3706598.3713363}, rather than manually scripting every plot branch and dialogue line.

Beyond interactive media, recent research began to explore authoring other forms of dynamic media that adapt to audience context as opposed to explicit interaction.
Kim et al. investigate creators’ perspectives on AI-mediated personalization in creative writing through interviews with authors considering hypothetical scenarios of AI-assisted narrative adaptation ~\cite{cla}. 
While participants recognized the potential for adaptive media to increase audience resonance, they emphasized mechanisms for preserving authorial intent.
Their findings highlight the importance of developing new authoring tools for adaptive media beyond interactive narratives—the primary focus of prior work—to other emerging forms of dynamic media such as adaptive music.

\begin{figure}[b]
  \centering
  \includegraphics[width=\linewidth]{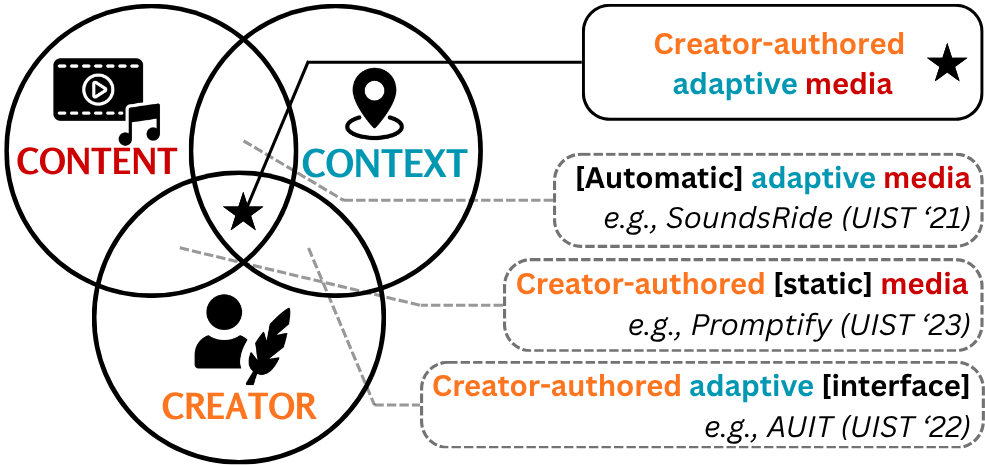}
  \caption{
  \approach unifies creator, content, and context to support creator-driven adaptive media.
}
  \label{fig:venn}
\end{figure}

\section{\approach for Adaptive Media}

\rv{We propose the \approach{} framework to support creators in authoring adaptive media.}

\rv{Unlike static media, adaptive media is not authored as a fixed artifact, but as behavior that unfolds across future consumption contexts. This changes the role of the creator from editing individual outputs to specifying how media should adapt while remaining faithful to artistic intent. C³ addresses this challenge by externalizing three core elements of adaptive authoring—creator, content, and context—as explicit representations that creators can inspect, author, and refine, as illustrated in \autoref{fig:venn}.}

\paragraph{Defining adaptive media}
We refer to adaptive media as content that responds to characteristics of the consumption context, drawing parallels from context-aware and ubiquitous computing~\cite{dey2001understanding, Lieberman2000context, Grubert2017context}.
Adaptive media could change based on the audience—their background, preferences, or current state; the environment in which it is consumed; tasks or activities performed by the audience during consumption; or the medium through which it is delivered.
This poses unprecedented challenges for creators, emphasizing the need for frameworks that enable meaningful artist control.

\paragraph{Authoring adaptive media}

Unlike static content, creators cannot directly review every possible outcome, and unlike interactive media, adaptive media responds implicitly to contextual factors, making it difficult for creators to predict or fully conceptualize all possible variations.
In interactive systems such as video games and narrative experiences, variation is driven by \textit{explicit} user inputs, allowing authors to define a bounded space of possible trajectories within the work \cite{10.1145/3706598.3713363}. 
In contrast, adaptive media adjusts \textit{implicitly} based on those contextual parameters as either a complementary or additional factor.
Contextual factors may differ substantially from the author’s own perspective and lived experiences.
Since creators need to define and consider context as additional input, we argue that this has implications for the whole workflow of creation.

\paragraph{A framework for adaptive media.}
Combining insights from interactive media authoring~\cite{1626183, green2019define, 10.1145/3706598.3713363} and context-aware computing~\cite{dey2001understanding, Lieberman2000context, Grubert2017context}, 
we define three design axioms (DAs) that capture the core characteristics of systems developed under our \approach{} framework.

\begin{itemize}
\item \textbf{DA1: Consider \textit{creator} intent.} 
Provide mechanisms for creators to articulate artistic goals to guide adaptations. 
\item \textbf{DA2: Consider \textit{content} structure.} 
Provide mechanisms for specifying which structural aspects must be preserved during adaptation.
\item \textbf{DA3: Consider \textit{context}-driven adaptations.} 
Provide mechanisms for shaping relevant audience context, and explore how media adapt for different contexts.
\end{itemize}

Following these goals, we conceptualize adaptive media authoring as the coordination of these three interdependent dimensions: \textit{creator}, \textit{content}, and \textit{context}. 
Rather than treating adaptation as a purely automated generation process, \approach exposes these dimensions as explicit elements of the authoring workflow. 
This allows creators to specify how adaptations should align with their artistic goals, structural constraints, and contextual signals.
For each dimension, we propose to consider a set of domain-agnostic inputs and steering actions.
We also provide domain-specific instantiations for the context of adaptive lyrics, which we implement in \projectname.
We see our approach as extensible within the individual categories, since each domain has specific considerations. 

\paragraph{\textbf{Creator}.} 
Adaptive media authoring considers the \textit{artistic intent} that guides how media adapts.
As \textit{inputs}, systems should consider parameters that capture the envisioned theme, narration voice and style, emotional message, and artistic vision.
Systems should \textit{steer and constrain} the space of how media is adaptive by considering \textit{where} those parameters are applied, for example globally enforcing a theme; \textit{how} those parameters are applied, for example how narration voice influences the media; which of those aspects are \textit{prioritized}, for example emphasizing narrative voice locally over a global theme to preserve a certain style. 
Finally, we define \textit{boundaries} of the adaptations, for example how far an adaptation can deviate from the original lyrics and the defined intent.

\paragraph{Creator considerations in adaptive lyrics}
Lyrics are semantically dense, where small changes can shift meaning, tone, or style. MultiVerse lets creators steer adaptations at three levels: global anchors (theme, emotion, style), local line guidance, and word-level locks. This explicit separation ensures AI-generated variations remain aligned with creator intent while allowing controlled flexibility.

\subsubsection*{\textbf{Content}.} 
Content captures the structural properties of a media artifact.
Different creative domains impose different structural requirements. 
For example, structural properties such as camera framing and the sequence of events are critical for narrative coherence in film editing~\cite{10.1145/3072959.3073653}.

In general, authoring tools need to enable creators to specify what aspects to \textit{preserve}, what parts can be \textit{relaxed/adapted}, and finally what can be \textit{added} or \textit{removed}.
Those actions can be defined on a global level (\eg whole characters can be added to a story) and tools need to enable creators to address downstream consequences.
Alternatively, content can be steered locally, \eg word- or sentence-level edits for text-based media, or color adjustments for image-based media.
Importantly, not all structural properties are equally important for every work. 
Creators therefore need to be able to specify which structural elements must be preserved and how adaptations should respect these constraints.

\paragraph{Content considerations in adaptive lyrics}
Lyrics consist of a rich set of structural properties, which have been explored in musicology~\cite{moylan2020recording} and music information retrieval~\cite{mayer2011musical}.
MultiVerse enables creators to steer adaptive lyrics by constraining factors such as syllable count, rhyme structure, stress patterns, and melodic alignment, ensuring that generated variations remain compatible with the underlying melody and overall song structure.

\subsubsection*{\textbf{Context}.} 
Context is the situational information that 
drives the adaptation of media artifacts, including aspects about end users (or audience), environment and system capabilities~\cite{Lieberman2000context}.
Each dimension has multiple parameters, such as an audience member’s preferences or prior experiences.
Adaptive authoring tools should allow creators to \textit{define} which contextual inputs are considered, specify their \textit{influence}, determine their \textit{priority}, and \textit{constrain} how contextual information guides adaptations.

\paragraph{Context considerations in adaptive lyrics}
Audience and situational context shape how lyrics are interpreted and experienced. 
While a song could, in principle, adapt to many contextual signals at the time of consumption (\eg listening location), in this work we focus on the audience as the primary driver of adaptation. 
MultiVerse enables the prototyping of this process through simulated listeners. 
Creators can proactively define which audience aspects are relevant, and then guide how adaptations are applied by linking those responses to specific parts of the lyrics.

\section{\projectname}

\begin{figure}[b]
  \centering
  \includegraphics[width=\linewidth]{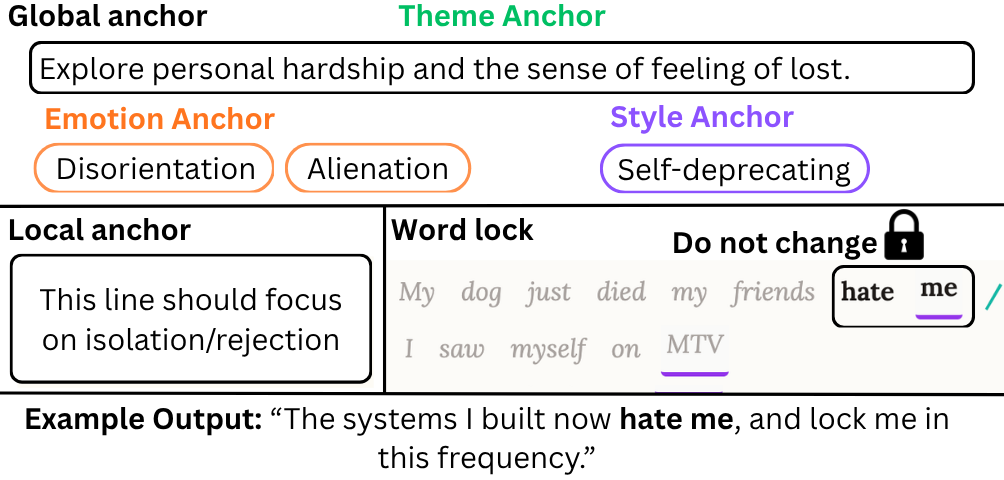}
  \caption{Creator intent authoring features in \projectname. Global anchors affect all generations, local anchors affect one line, and word locks keep specific words unchanged.
}
  \label{fig:creator}
\end{figure}

To explore \approach in practice, we instantiate our approach in the domain of songwriting through the design of \projectname, a system that enables creators to steer adaptive lyrics by authoring steering controls and testing adaptation behavior with simulated personas.

Concretely, \projectname takes a finished set of original lyrics as \textit{input}. Creators then author steering controls that specify how the lyrics should adapt across different audience contexts. 
Using simulated personas, creators preview potential adaptations to explore variations and iteratively refine their controls. 
The \textit{output} of this process is a set of steering controls that guide how the lyrics adapt.
In this section, we describe the design and implementation of \projectname, organized around the design goals: features that support the specification of creator intent (\ref{creator}), content structure (\ref{content}), and adaptation context (\ref{context}). 

\subsection{Creator: authoring artistic intent}\label{creator}
\projectname provides multi-level control for steering adaptive content, supporting both high-level guidance of intent and low-level control over line-level adaptations (Fig \ref{fig:creator}).

\paragraph{Global intent anchors.} \projectname supports the alignment of high-level artistic intent through specifying a set of global anchors based on 3 dimensions: 
\begin{itemize}
    \item Theme: the core message and concept of the song.
    \item Emotion: the target emotion to evoke.
    \item Style: the voice and aesthetics of the lyrics agnostic to theme and emotion.
\end{itemize}

The lyrics are automatically analyzed to generate an initial set of tags, while also supporting iterative refinement of this document with the creator. This anchor document is provided to the prompt for all lyric generation tasks to steer the output. Emotion and style anchors are short keyword tags, while the theme anchor is a more elaborate statement in natural language. 

\paragraph{Local intent anchors.} \projectname additionally supports the optional authoring of short intent statements to steer the behavior of each individual line. 

\paragraph{Word locks.} \projectname allows creators to lock or unlock individual words, ensuring they remain unchanged during adaptive generation.

\subsection{Content: authoring music structure}\label{content}
\begin{figure}[t]
  \centering
  \includegraphics[width=.8\linewidth]{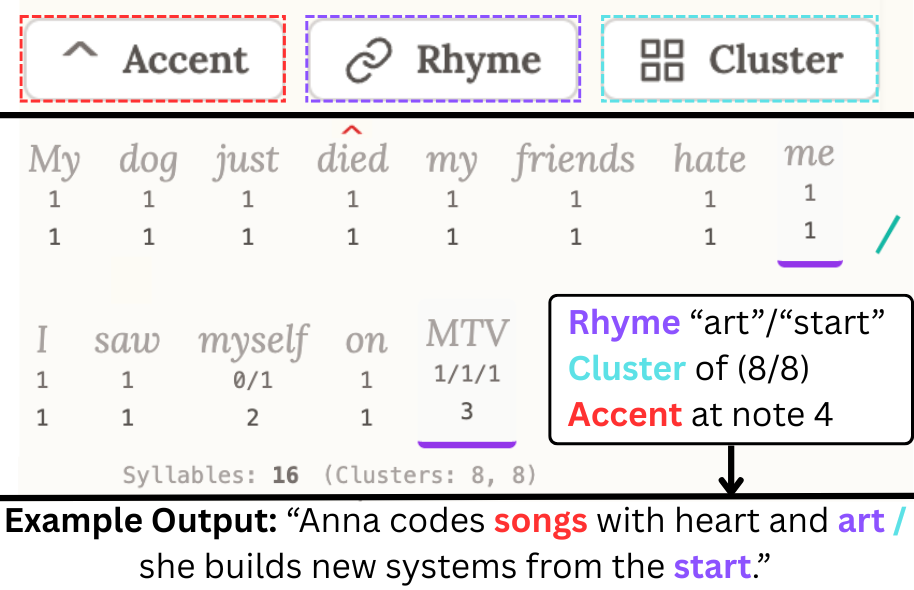}
  \caption{Content structure authoring features in \projectname. 
  Accent ensures the labeled syllable is stressed, Rhyme ensures selected words rhyme, and Cluster prevents multisyllabic words from crossing labeled boundaries.
}
  \label{fig:content}
\end{figure}

\projectname provides tools for controlling structure (Figure \ref{fig:content}).

\paragraph{Syllable count.} When writing lyrics to a predetermined melody, syllable count must match the number of vocal notes, as each note typically carries one syllable. \projectname enforces this as a hard constraint to preserve alignment with the fixed melodic structure. 

\paragraph{Melody clustering.} Even when the line-level syllable count is correct, word-level coherence can still be disrupted if a word is split across discrete melodic segments. 
For example, when the lyrics "misleading list" is sung with a pause in the middle, it may sound like "miss Lee (pause) dean list" and confuse the listener.
To prevent this, \projectname allows creators to label where breaks occur. The system then ensures that the number of syllables in each cluster matches the corresponding notes.

\paragraph{Stress pattern.} 
Beyond syllable count, natural-sounding lyrics require alignment between lyrical stress and melodic emphasis. In many melodies, certain notes fall on strong beats or receive emphasis through duration or pitch. Placing an unstressed syllable in these positions can sound awkward or alter the meaning. For example, the word ``present'' can be pronounced as \textit{PRE-sent} (a gift) or \textit{pre-SENT} (to demonstrate).
To prevent this, \projectname allows creators to mark syllable positions that should be lexically stressed.

\paragraph{Rhyme.} Rhyme plays a key role in lyrical structure and listener expectation. \projectname allows users to label pairs of words as rhyming constraints, ensuring that any subsequently generated lyric variation preserves the specified rhyme relationships.

\begin{figure}[t]
  \centering
  \includegraphics[width=\linewidth]{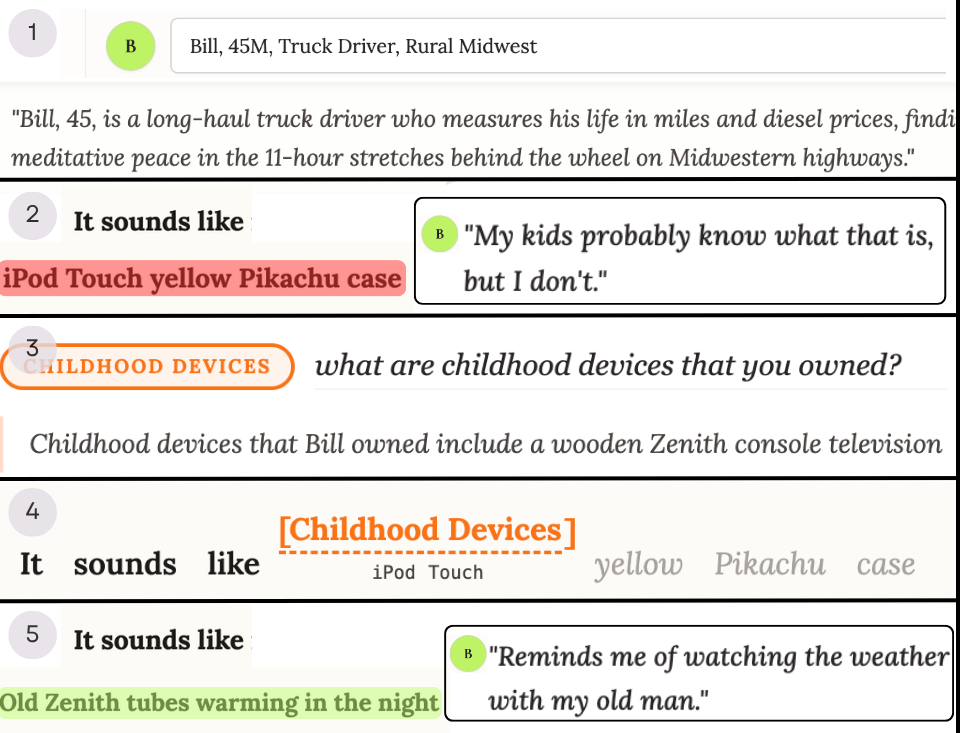}
  \caption{
  \rv{
  Persona-based authoring in MultiVerse. Creators query listener context, bind responses to lyric variables, and preview personalized lyric adaptations.}
  \vspace{-1em}
}
  \label{fig:persona}
\end{figure}

\subsection{Context: Audience-aware authoring}\label{context}
\projectname enables audience-aware authoring through simulated personas, allowing creators to preview feedback and define relevant audience context to shape system behavior (Figure \ref{fig:persona}).

\paragraph{Listener personas.}
\projectname includes a set of fictional listener personas that represent diverse audience contexts. Each persona is described through nine profile dimensions capturing identity, background, and lifestyle factors that may shape how a listener interprets a song. These dimensions include name, demographics, location and cultural context, occupation, background and formative experiences, social relations, physical appearance or aesthetic style, personality traits, and daily routines.

\paragraph{Contextual query and variables.}
While personas provide a rich contextual profile, they cannot capture every aspect of a listener’s experiences. In principle, an audience member’s background could be described through an arbitrarily detailed biography, which would be impractical to represent within the system or incorporate into generation prompts. 

To address this limitation, \projectname allows creators to proactively define the aspects of audience context that are most relevant to their song through \emph{contextual queries}. A contextual query is a question posed to the listener persona that elicits information aligned with the song’s themes or narrative. The response can then be linked to \emph{variables} that replace specific text in the lyrics. 

\paragraph{Listener feedback. }The system simulates fictional audience personas, showing their contexts and reactions to different lyrics. By observing these simulated responses, creators can adjust controls to better align adaptations with intended audience experiences.

\begin{figure*}[t]
  \centering
  \includegraphics[width=\textwidth]{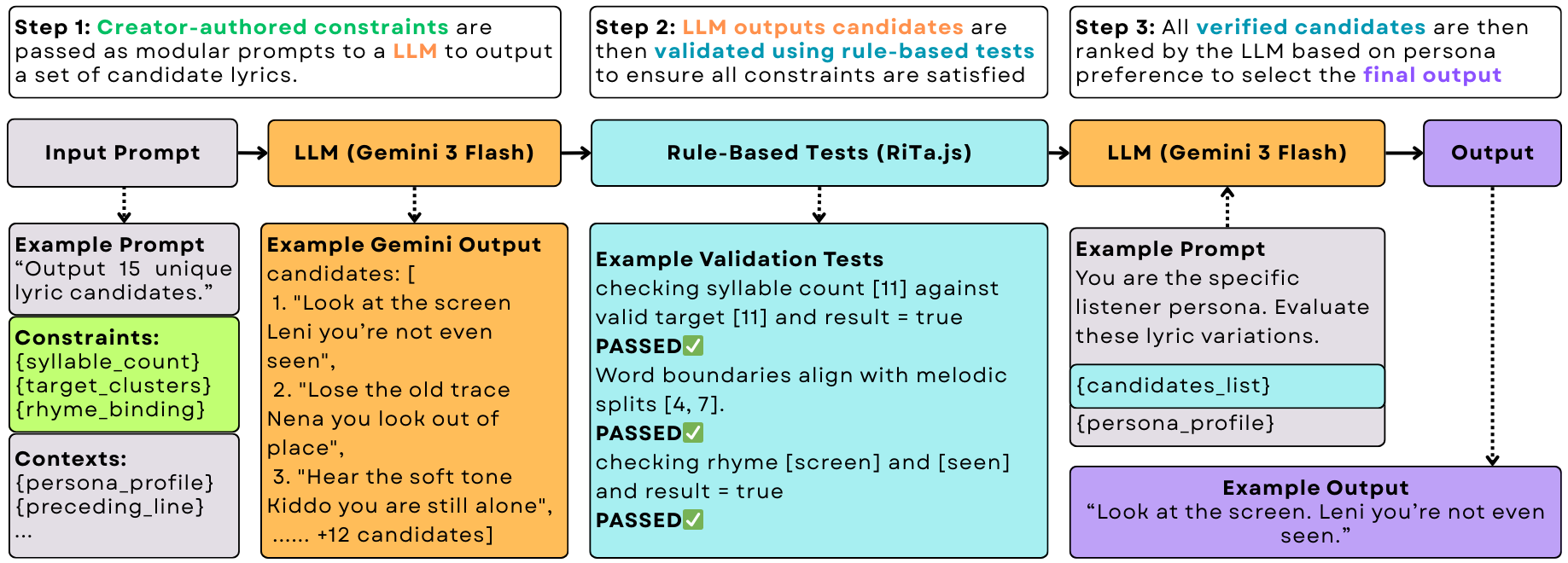}
  \caption{\textbf{Lyric generation pipeline:} Creator-authored constraints are passed as modular prompts to the LLM to generate an initial set of candidates. Each candidate is then validated using rule-based tests to ensure all constraints are satisfied. All verified candidates that pass the tests are then ranked by the LLM based on persona preference to select the final output.}
  \label{fig:pipeline}
\end{figure*}

\subsection{Implementation}

The interface of \projectname is implemented as a web application using React. Adaptive lyric generation is performed through API calls to Gemini 3 Flash, and then validated with rule-based computational linguistic library RiTa.js \cite{RiTa2025}.

\paragraph{Preprocessing and control initialization.}
\rv{When a song is loaded, MultiVerse initializes creator controls by using RiTa.js to infer structural properties (e.g., syllable count, stress, and rhyme). These automatically populated controls serve as editable defaults that creators can review and override. The system also derives an initial creative vision (theme, emotion, and style) from the source lyrics.}

\paragraph{Adaptive lyric generation.}
To generate lyrics, \projectname prompts Gemini using user-authored controls together with persona information (Fig \ref{fig:pipeline}). Because the generated lyrics must satisfy strict structural constraints (e.g., syllable count, rhyme bindings), generation is performed on a per-line basis, 
with the prior line available in context.
This allows the system to validate each candidate line deterministically before proceeding. The prompt is constructed from modular components representing creator-defined constraints, such as locked words, variable spans, rhyme bindings, intended line meaning, listener context, and song-level creative goals.

\paragraph{Batch generation. } Rather than generating a single output line at a time, the system requests 15 candidate lyric lines in each attempt. This provides a pool of alternatives that can be filtered and ranked.

\paragraph{Prompt variation.} The system also cycles through several prompt formulations. The default prompt asks the model to refer to the structure of the original line, while a second prompt asks the model to ignore the original line. A third prompt provides a partially complete template and asks the model to complete the lyric using the provided fragments. If a line contains locked words or variables, the template-completion prompt is prioritized. 

\paragraph{Constraint validation.}
Each batch of generated candidates is filtered using deterministic validation checks. For verifiable constraints---including syllable counts, phrase-cluster boundaries, locked-word preservation, accent placement, and rhyme matching---\projectname applies rule-based checks implemented with RiTa.js. 
If no candidate in a batch satisfies the constraints, the system proceeds to the next prompt variation. Otherwise, the valid candidates from the first successful batch are retained for further evaluation. 

\paragraph{Candidate ranking.}
After validation, the remaining candidates are evaluated with a second LLM call. The model assesses them from the listener’s perspective and ranks them by expected preference, and the top-ranked candidate is selected as the final output.

\section{User Study}

We conducted a user study with songwriters to explore how creators approach the task of steering adaptive lyrics and how different workflows shape this process. As there is no established method for this task, our goal was not to evaluate performance against a canonical approach, but to use contrasting workflows to surface trade-offs in how this emerging form of media can be steered. We examine how creators reason about encoding adaptation into their songs, what trade-offs arise, and how these workflows affect their sense of control and authorship.

\paragraph{Workflows}
Participants used their own songs to create adaptive lyrics using two contrasting workflows: \textbf{\projectname}, and  \textbf{\baseline}, a prompt-based workflow that relies on free-form text instructions. 
As there is no established method for steering adaptive lyrics, the prompt interface represents a hypothetical approach that is similar to everyday AI chat tools. 
Relative to the structured approach of \projectname, \baseline{} is more ``hands off''---all creator steering is conveyed through a text box and executed by an LLM (Gemini).
The prompt-based workflow features an interface consisting of a persona selector identical to full \projectname, windows to view original lyrics and generated lyrics, and a text box to input the adaptation instructions. 
Unlike AI chatbots, \baseline does not allow back-and-forth conversation, as no artist iteration is allowed during runtime adaptation. Instead, both workflows involve a prompt template requesting Gemini to alter \{\textit{original lyrics}\} based on \{i\textit{ntended audience}\} while adhering to \{\textit{control instructions}\}.
The adaptation instruction is applied to the entire set of lyrics rather than individual lines, and since the instructions are in free-form text, no rule-based validations can ensure they are followed. 

\subsection{Participants}

We recruited 10 skilled participants (7 male, 3 female; age: $M = 23.7$, $SD = 3.47$) with at least 2 years of songwriting experience and at least 2 original compositions. Participants had diverse musical backgrounds, and represented both music students and professional musicians. On average, participants were highly skilled and reported $M = 9.2$ years ($SD = 6.14$) of songwriting experience and $M = 8.2$ years ($SD = 4.64$) of formal musical training. Several participants had professional experience composing, producing, or performing music, including releasing albums, composing for games and films, performing live, and teaching music. Table~\ref{tab:participants} in the appendix provides details of participant backgrounds. Participants were recruited through online outreach and convenience sampling.

\subsection{Procedure} 

Participants provided informed consent, demographic information, and were introduced to the study goals and the concept of AI adaptive lyrics. They then completed two conditions in counterbalanced order: \projectname, and \baseline.
Each condition followed the same structure. Participants first completed a brief guided tutorial on the interface, including a walkthrough exercise for creating adaptive lyrics for a provided song (approximately 5 minutes for the prompt condition and 10–20 minutes for \projectname).
They then performed a 20-minute authoring task in which they adapted one of their own songs (collected prior to the study and truncated to 8--16 lines) to support different audiences while preserving its original intent. In the \projectname condition, participants used the full interface to specify controls. 
In the \baseline condition, they authored instructions in natural language. 
In both cases, lyrics were generated using Gemini 3 Flash based on the original song, participant-authored controls, and target audience personas. 
All participants used the same 12 personas (four 3-persona sets).
For each condition, one set of personas was used for authoring, and another set for evaluation, \ie participants reviewed the outputs for a set of unseen personas to get a sense of how their adaptive lyrics would be used.

After each condition, participants completed a post-task questionnaire (7-point Likert scale) assessing their perceptions of the system and the resulting outputs. This includes 5 dimensions of the creativity support index \cite{10.1145/2617588} and 3 modified items from the mixed-initiative creativity support index \cite{lawton2023drawing}.

After completing both conditions, they took part in a semi-structured interview reflecting on their experiences with the two workflows and broader perspectives on authoring adaptive music. Interviews were audio-recorded, transcribed, and analyzed through affinity diagramming. We logged all participant-authored control interactions.

\section{Results}
\rv{We organize our results into two parts. We first present comparative findings from contrasting MultiVerse with prompting-based authoring, highlighting the strengths, limitations, and design trade-offs of different workflows. We then present broader creator perspectives on adaptive media, including why creators value adaptive media, how it reshapes the creative process, and how it influences notions of authorship.}
\rv{\subsection{Comparative insights}}
\rv{Overall, both workflows were positively received (\eg ``\textit{I enjoyed using the system}'' was rated $M = 5$ for both), with no statistically significant differences across subjective ratings (all ${p > .05}$, Table \ref{tab:survey}). Rather than indicating a clear preference for one workflow, participants described complementary strengths and trade-offs between the two approaches. Participants consistently appreciated the control and context exploration mechanisms afforded by MultiVerse, while also recognizing that the control introduced trade-offs in adaptation flexibility and iteration speed.}
\subsubsection{\rv{Context-based authoring mechanisms enabled creators to steer adaptations using meaningful context}} 

\rv{Rather than adapting lyrics without regard for contextual relevance, participants wanted to specify contexts they considered meaningful for the song itself. As P4 explained, the system often by default "focuses on profession, but that's not what I want to optimize for." MultiVerse's contextual querying mechanism allowed creators to proactively define which aspects of a listener's context should drive adaptation. Nearly all participants (9/10) proactively queried contextual information and incorporated the information into their lyrics, most commonly asking questions aligned with their songs' themes (6/10). For example, participants writing love songs asked about breakup details and regrets in love (P3, P5), while more introspective songs prompted questions about recent experiences relevant to the song's emotional themes, such as recent regrets (P10). Participants also tailored queries to the structure of their lyrics (9/10), requesting contextual details that could be directly mapped into lyrical elements, such as sounds heard during work (P1), places where listeners reflect alone (P2), daily activities and environments (P8), current headspace and imagery (P9), or everyday habits such as where listeners get their news (P10). Participants described these creator-defined adaptation contexts as producing lyrics that felt more genuine and less "surface-level" than adaptations based only on generic audience attributes (P9).
}
\subsubsection{\rv{Structured controls gave creators greater confidence over adaptation outcomes.}} 
Participants consistently associated MultiVerse with greater control over lyrics adaptation when compared to \baseline (8/10).  
Participants attributed this perception primarily to the ability to define explicit constraints on how lyrics could change, including guarantees that specific lines would remain unchanged. 
As P1 explained, \projectname “allows you to specify which parts stay original and which parts change, so it feels safer and more controlled.”
In contrast, the simplified workflow was frequently described as less consistent and predictable. Because instructions were embedded in free-form prompts, participants often had to “guess about how much of the prompt the AI would follow,” with the system sometimes ignoring parts of their instructions (P4). The resulting unpredictability also raised concerns about consistency across generations. As P10 explained, if the system were used in practice, they would be “nervous about generating the adaptation again and not liking the next version—even though I haven’t changed anything (in the control prompt).”

\subsubsection{\rv{Strict constraints reduced adaptation flexibility and iteration speed.}}
\rv{Participants observed trade-offs in lyric quality with strict constraint enforcement (7/10). While explicit constraints over accent, rhyme, and clusters improved structural consistency (P1), several participants felt they also narrowed the adaptation space. As P9 explained, constrained edits sometimes produced fragmented lyrics where "some of the lyrics were mine, some were generated, and some were edited," whereas the prompt-based workflow could better fit a listener's story, even if it diverged from the original intent. Participants also emphasized the importance of rapid iteration for refining adaptive lyrics (7/10). While the Prompt workflow supported rapid exploration through full-song generations ("the faster it is, the faster I can say whether I like it," P5), MultiVerse's more granular authoring workflow slowed iteration, leaving several participants feeling they had insufficient time to fully refine their controls during the study session.}
\rv{\subsection{Creator perspectives on adaptive media}
Beyond comparing authoring workflows, participants reflected on adaptive lyrics as an emerging creative medium and how it differed from conventional songwriting. 
Our interviews revealed perspectives that extended beyond the specific interfaces, including why they found adaptive media valuable, how authoring adaptive media introduces a distinct creative process and new structural considerations for composition, and how it reshapes notions of collaboration, authorship, and ownership.}

\subsubsection{\rv{Creators valued adaptive media for enabling new forms of audience connection}}
Artists valued the personalization of adaptive lyrics for its potential to strengthen emotional connection with listeners.
Study participants noted that adapting lyrics for specific listeners could make songs more meaningful and relatable, and help them achieve common goals of songwriting such as helping listeners ``feel something'' (P10) or providing experiences that are "really meaningful" and even "healing" (P4).
\rv{Participants described adaptive lyrics as enabling fundamentally new relationships between artists and audiences. P9 envisioned that instead of searching for songs that happened to resonate with their lives, superfans could "customize lyrics to their experiences," allowing their favorite artists to sing their stories. P4 similarly envisioned adaptive music becoming a new form of fan participation where fans co-create personalized tracks with their favorite artists, extending existing parasocial experiences such as K-pop personal chats that personalize artist-authored messages to create the illusion of one-to-one interaction \cite{yang2022kpop}. P1 framed adaptive lyrics as a new form of lyrical "remixing," where audience-specific details could "resonate so hard" by reflecting listeners' own lived experiences.}

Participants, however, also raised concerns about too much personalization. Tailoring lyrics too closely to a specific listener could reduce interpretive openness, as songs can “take on different meanings” (P6).
Furthermore, lyrics may become “too hyper-tailored” to feel broadly relatable (P3), as well as lead to decreased perceived authenticity that feels ``kind of fake'' (P10).

\subsubsection{\rv{Authoring adaptive media is a distinct creative process}}
\rv{
Participants described adaptive lyric authoring as an iterative process of discovering which aspects of a song should remain invariant across future adaptations. Rather than treating artistic intent as something fully specified before authoring, they refined it through repeated cycles of generating adaptations, observing what changes challenged their intent, and updating their steering accordingly. As P6 explained, adaptive authoring required them to "deconstruct the lyrics," and decide "which lines can be variable and which lines need to stay the same." P7 similarly described the process as "a feedback loop" of "understanding what's important in the lyrics" and "what about the lyrics is important to the artist," then "using that information to generate more content" to continue the loop. Participants noted that this process also helped them better understand their own songwriting. P6 explained that this iterative process "made it very clear to me what the song was really about" and helped them "understand the mechanics of my lyrics much better."}

\subsubsection{\rv{Adaptive media favored different content structures}}

\rv{Participants distinguished between songs that naturally lent themselves to adaptation and those that did not. P6 observed that some songs are "more adaptable than others," explaining that lyrics referencing specific places naturally support variation across audiences. Similarly, P4 contrasted two of their own songs, noting that one built around abstract metaphors could benefit from adaptation because personalization might help listeners better understand its intended meaning, whereas another centered on seasons was already broadly relatable and required little personalization. In contrast, participants viewed deeply personal narratives as poor candidates for adaptation. As P9 explained, adapting a song about losing a parent could feel inappropriate because "it's not your story to tell."}

\rv{This became particularly apparent when participants adapted finished songs. P10 explained that their songs were "already written" and "finalized in my head as the lyrics," making it difficult to rethink how the song could vary. Similarly, P8 noted that their process would "look a lot different if I were starting from scratch." Rather than composing freely, adaptation became "identifying certain phrases or words that might make sense to swap out" because the existing structure and rhyme scheme made the process "a lot more restricted."
Reflecting on the experience, both participants suggested they would compose adaptive songs differently from the outset, designing lyrical structures that better accommodate future variation.}

\subsubsection{\rv{Adaptive media reshaped notions of collaboration, authorship, and ownership}}

\paragraph{Collaboration}
Participants framed the relationship as hierarchical, positioning themselves as responsible for the overall creative vision while the AI carried out specific tasks. 
Several used managerial metaphors, describing themselves as a “CEO” (P4) or “creative director” (P9) directing an assistant who handles creative details based on their instructions, sometimes reduced to a “word chooser” (P8).
However, collaboration broke down when the system appeared to act independently of the creator’s intent. 
P10 described the process as frustrating when the AI ignored explicit instructions, saying that in those moments the system had “more of a say in the final outcome” than they did, while P7 remarked that simply clicking “generate” left them feeling “out of the loop.” As P3 summarized, the tool was “supposed to be a collaboration,” but at times felt like an argument with a collaborator who wanted to “do its own thing.”

\paragraph{Authorship}
Substantial modifications often reduced artists’ sense of authorship and led to reluctance in delegating the writing process to AI.
Because songwriting typically involves careful word choice, even small changes could make the result feel misaligned with the artist’s voice. 
\rv{For example, P2 explained that in songwriting “we choose every word very carefully,” so even changing “just one word feels like it's a little off.” }
When the generated lyrics diverged further from the original song, participants sometimes felt disconnected from the outputs, or lost the sense of authorship. 
P10 remarked that the output was “very quickly becoming not my song, because it was using words that I didn't even know.” 
This highlights the need to balance newly generated output with creator-specified constraints.

\paragraph{Ownership}
Participants expressed various views about who owns AI-adapted lyrics. 

\rv{P1 (AI engineer), who already uses AI in their creative workflows, maintained full ownership, viewing AI as a tool rather than a co-author.}

\rv{Others described ownership as distributed across multiple actors shared between "me (the artist), the listener, and the AI" (P6). 
For instance, P4 suggested that ownership may lie with the listener, as their inputs shape the adaptation into an “audience-authored” version of the song.
Similarly, participants described sharing ownership with the algorithm (P10), or the human developer (P7) and company (P8) who created the AI. 
Finally, some participants framed AI-generated adaptations as derivative works inspired by the creator’s style rather than authored by them. }
P9 compared the process to AI-generated visual art that is “Monet-inspired”—reflecting an artist’s influence without actually belonging to them—and argued that while the output might not fully belong to the songwriter, creators should still receive credit and compensation for work derived from their artistic style.
Furthermore, incorporating audience information makes the adaptation reflect the listener’s persona, creating a sense that the song becomes “audience-authored” (P4), thereby introducing greater ambiguity in authorship and ownership.

\section{Discussion}

\rv{We discuss limitations of our approach, and outline directions for future research.}
\rv{\subsection{Content adaptation beyond lyrics}}

\rv{We instantiated our adaptive media authoring framework in the domain of songwriting. However, we anticipate advances in generative AI may enable other forms of adaptive multimedia. We hope to explore these richer forms of adaptive media and how the \approach framework can be instantiated in different domains.}

\paragraph{Authoring adaptive songs}
MultiVerse currently focuses on text (\ie lyrics) and assumes a fixed melody with finalized lyrics. Future systems could allow creators to compose multiple melodic variations or entirely new tunes alongside lyrics. Supporting this would require tools for generating, adapting, and previewing musical content, while giving creators control over which melodic elements are preserved, modified, or recombined.

\paragraph{Authoring adaptive media beyond music}
While we demonstrate our approach in songwriting, other media—such as film, audiobooks, and poetry—could also be adaptive. For example, comics could vary dialogue or visuals, and podcasts could adjust tone for different listeners.
In such settings, creators face the same core challenge: specifying how content should change across contexts without directly reviewing each variation. This suggests that steering may be a useful abstraction for adaptive media more broadly, and \approach offers a foundation for navigating this emerging design space.

\rv{\subsection{Context acquisition and deployment}
Adaptive media ultimately depends on both creator and audience adoption. While this paper focuses on how creators author adaptive experiences, an equally important direction for future work is understanding how audience context is acquired in deployed systems and how listeners respond to context-aware adaptations. }
\rv{\paragraph{Envisioned deployment} Deployment requires deciding what audience context to use, whether that context should trigger personalization at all, and how personalized content should be presented to listeners. Audience context may be explicitly provided by listeners (\eg an extended version of a user profile) or inferred from interaction history, sensors, or user models \cite{shaikh2025creating}. Context availability, however, does not imply that personalization should always occur. Adaptive systems may instead reserve personalization for moments when context meaningfully complements the creator’s intent or the listener’s situation, and otherwise present the original work.}
\rv{\paragraph{Audience perspectives} 
Deployment decisions also raise questions of privacy and user agency. Future work may draw inspiration from work on privacy-aware adaptation techniques and privacy negotiation in XR to develop adaptive media systems that balance personalization with audience privacy preferences \cite{privacyadaptation, privacyEquilibrium}. More broadly, adaptive media should be evaluated not only by what context it can use, but by how listeners experience those choices: whether personalization feels meaningful, appropriate, intrusive, or authentically tied to the work. Future work should therefore examine which deployments listeners perceive as worthwhile, which contextual cues they find acceptable, and when the original work should remain unchanged.}

\rv{\subsection{Creator challenges for authoring adaptive media}
Unlike static media, adaptive media is experienced through many runtime-generated variations. Because creators cannot directly review every possible adaptation, authoring becomes the task of specifying behavior over a potentially unbounded space of future outputs. This introduces a fundamental challenge for adaptive media: how can creators meaningfully control outputs they will never directly observe? }
\rv{\paragraph{From steering to explicit adaptation policies} One approach to addressing this challenge is to explore authoring explicit and verifiable adaptation policies. Most current AI-assisted creative tools rely on natural-language prompts to influence model behavior, providing limited guarantees that generated outputs will consistently satisfy creator intent. This becomes particularly problematic for adaptive media, where outputs are generated at runtime and cannot be exhaustively reviewed. MultiVerse explores one step in this direction by combining generative models with deterministic verification of creator-authored constraints. However, our study also revealed that not all constraints should be treated equally. For example, one songwriter observed that preserving a rhyme constraint forced the system toward lexically compatible words that conflicted with the intended emotional tone of the lyrics. This suggests that future authoring tools should allow creators to express different levels of commitment to adaptation policies, specifying which requirements must always hold, which are preferred but negotiable, and how conflicts among competing policies should be resolved. Supporting explicit prioritization and conflict management may become increasingly important as adaptive media grows in complexity and the space of possible adaptations expands.}
\rv{\paragraph{Modeling the author} While richer adaptation policies can provide stronger guarantees over runtime behavior, manually specifying them may become increasingly burdensome as adaptive experiences grow in complexity. Rather than explicitly authoring every adaptation policy, creators may instead author a computational representation capable of making low-level creative decisions during runtime. During our study, some participants wanted to provide the system with much richer representations of their creative process—for example, conversations, drafts, or other materials that inspired a song—to better preserve artistic intent across future adaptations. One way to realize this vision is to treat authoring as a longitudinal process, allowing systems to gradually construct richer representations of artistic intent from interaction traces such as drafts, prior prompts, and refinement decisions (e.g., which variations are accepted, rejected, or edited)~\cite{hammad2025towards, smith2025fuzzy, smith2025scaling, lin2021learning}. Similar to recent work on conversational agents that act on a user's behalf~\cite{cheng2025conversational}, such representations could enable creators to delegate creative decisions to the system while remaining aligned with their goals. Under this view, the published artifact is no longer only the work itself, but also the computational representation that continues shaping how the work evolves across audiences and contexts.}

\section{Conclusion}
We presented \textit{\approach}, a creator-centered approach for steering adaptive media, and instantiated it in \projectname for adaptive lyric authoring.
As dynamic media experiences expand beyond interactivity, new authoring paradigms will be needed to support creators in designing how their work responds to diverse consumption contexts. 
Through MultiVerse, we show that domain-specific tools can help creators steer adaptations toward outcomes that align with their artistic intent.
More broadly, we hope our work encourages further exploration of creator-centered tools for context-adaptive media across domains such as music, video, and other emerging forms of dynamic media.

\bibliographystyle{ACM-Reference-Format}
\bibliography{sample-base}

\clearpage
\appendix

\begin{table*}[t]
\centering
\caption{Participant musical backgrounds. FT = years of formal musical training; SW = years of songwriting experience.}
\label{tab:participants}

\begin{tabular}{c c c p{9cm} p{4cm}}
\toprule
ID & FT & SW & Musical Practice & Genres \\
\midrule

P1 & 10 & 5 & EDM producer; pianist, drummer, and DJ; experience with AI-assisted songwriting & Progressive/Future House, EDM, Pop/Punk \\

P2 & 16 & 10 & Graduate screen scoring student; composer for indie games and films; released album; active songwriter across genres & Electronic Pop, Alt Pop, Indie \\

P3 & 1 & 4 & Vocal training; hobby songwriter and producer & Pop, R\&B \\

P4 & 14 & 5 & Violinist with orchestral and jazz background; writes instrumentals and lyric adaptations & Pop \\

P5 & 0 & 5 & Collegiate a cappella performer; co-wrote and produced albums for collaborators & Bedroom Jazz Fusion \\

P6 & 4 & 16 & Primary songwriter and lyricist for two bands; producer, engineer, and performer & Indie Rock, Synth-pop, Electronic, Experimental \\

P7 & 10 & 15 & Songwriting and production for multimedia projects (theater, animation) & Indie \\

P8 & 6 & 5 & Live performer; released original songs; narrative-focused lyric songwriter & Folk Rock \\

P9 & 16 & 5 & Electronic music major; classical pianist; commissioned composer working professionally in music and film & Pop, R\&B, Alternative, Cinematic \\

P10 & 15 & 12 & Audio engineering and music production professor at university; active songwriter (50--100 songs); multi-instrumentalist and music educator & Pop, Rock, Indie \\

\bottomrule
\end{tabular}

\end{table*}

\begin{table*}[t]
\centering
\begin{tabular}{p{0.6\linewidth}cc}
\toprule
\textbf{Question} & \textbf{Mean MultiVerse} & \textbf{Mean \baseline} \\
\midrule
I would be happy to use this system or tool on a regular basis. & 4.5 (SD = 1.3) & 4.1 (SD = 2.1) \\
I enjoyed using the system or tool. & 5.0 (SD = 1.3) & 5.0 (SD = 1.6) \\
It was easy for me to explore many different ideas, options, designs, or outcomes using this system or tool. & 5.2 (SD = 1.6) & 4.6 (SD = 1.8) \\
The system or tool was helpful in allowing me to track different ideas, outcomes, or possibilities. & 4.8 (SD = 1.5) & 3.6 (SD = 1.8) \\
I was able to be very creative while doing the activity inside this system or tool. & 4.5 (SD = 1.6) & 4.8 (SD = 1.7) \\
The system or tool allowed me to be very expressive. & 4.2 (SD = 1.8) & 3.9 (SD = 1.6) \\
My attention was fully tuned to the activity, and I forgot about the system or tool that I was using. & 4.1 (SD = 1.9) & 3.9 (SD = 2.1) \\
I became so absorbed in the activity that I forgot about the system or tool that I was using. & 3.5 (SD = 1.8) & 3.4 (SD = 1.8) \\
I was satisfied with what I got out of the system or tool. & 4.0 (SD = 1.8) & 4.4 (SD = 1.6) \\
What I was able to produce was worth the effort I had to exert to produce it. & 4.1 (SD = 1.7) & 4.5 (SD = 1.4) \\
I was able to effectively communicate what I wanted to the system. & 4.3 (SD = 1.3) & 4.3 (SD = 1.7) \\
I was able to steer the system towards output that was aligned with my goals. & 4.7 (SD = 2.1) & 5.0 (SD = 1.7) \\
At times, I felt that the system was steering me towards its own goals. & 4.5 (SD = 1.7) & 4.6 (SD = 1.7) \\
I felt the lyrics created were mine. & 2.9 (SD = 1.8) & 2.8 (SD = 1.3) \\
The music created was totally due to my contributions (as opposed to the system's contributions). & 3.3 (SD = 1.5) & 3.8 (SD = 1.4) \\
I felt like I was collaborating with the system. & 4.7 (SD = 2.1) & 5.1 (SD = 1.7) \\
I believe the generated lyrics fits well with the original melody. & 4.2 (SD = 1.8) & 4.5 (SD = 2.0) \\
I believe the generated lyrics preserved the original intent of the song. & 4.0 (SD = 1.7) & 3.8 (SD = 1.8) \\
I believe the generated lyrics adapted appropriately for different audiences. & 4.9 (SD = 1.2) & 4.5 (SD = 1.6) \\
I believe the generated lyrics are of high quality. & 3.3 (SD = 1.6) & 4.0 (SD = 1.3) \\
I am satisfied with the generated lyrics overall. & 3.4 (SD = 1.9) & 3.8 (SD = 1.9) \\
I felt that the system helped me make informed decisions during the authoring process. & 4.8 (SD = 1.4) & 3.6 (SD = 1.9) \\
I could reasonably predict how the system would adapt the content after I finished authoring it. & 4.8 (SD = 1.5) & 3.8 (SD = 1.8) \\
I felt I had meaningful control over how the lyrics would adapt to different audiences. & 4.9 (SD = 1.6) & 3.9 (SD = 1.6) \\
I clearly understood how my controls affected the system’s adaptive behavior. & 5.0 (SD = 1.2) & 4.0 (SD = 1.8) \\
I trust the system to generate acceptable versions of the content that I might never directly review. & 3.0 (SD = 1.5) & 2.7 (SD = 1.6) \\
I believe the system can effectively generate adaptations for a large number of different contexts. & 4.5 (SD = 1.4) & 3.5 (SD = 2.1) \\
\bottomrule
\end{tabular}
\caption{Survey questions and mean ratings for MultiVerse and prompting.}
\label{tab:survey}
\end{table*}

\begin{figure*}
    \centering
    \includegraphics[width=\textwidth]{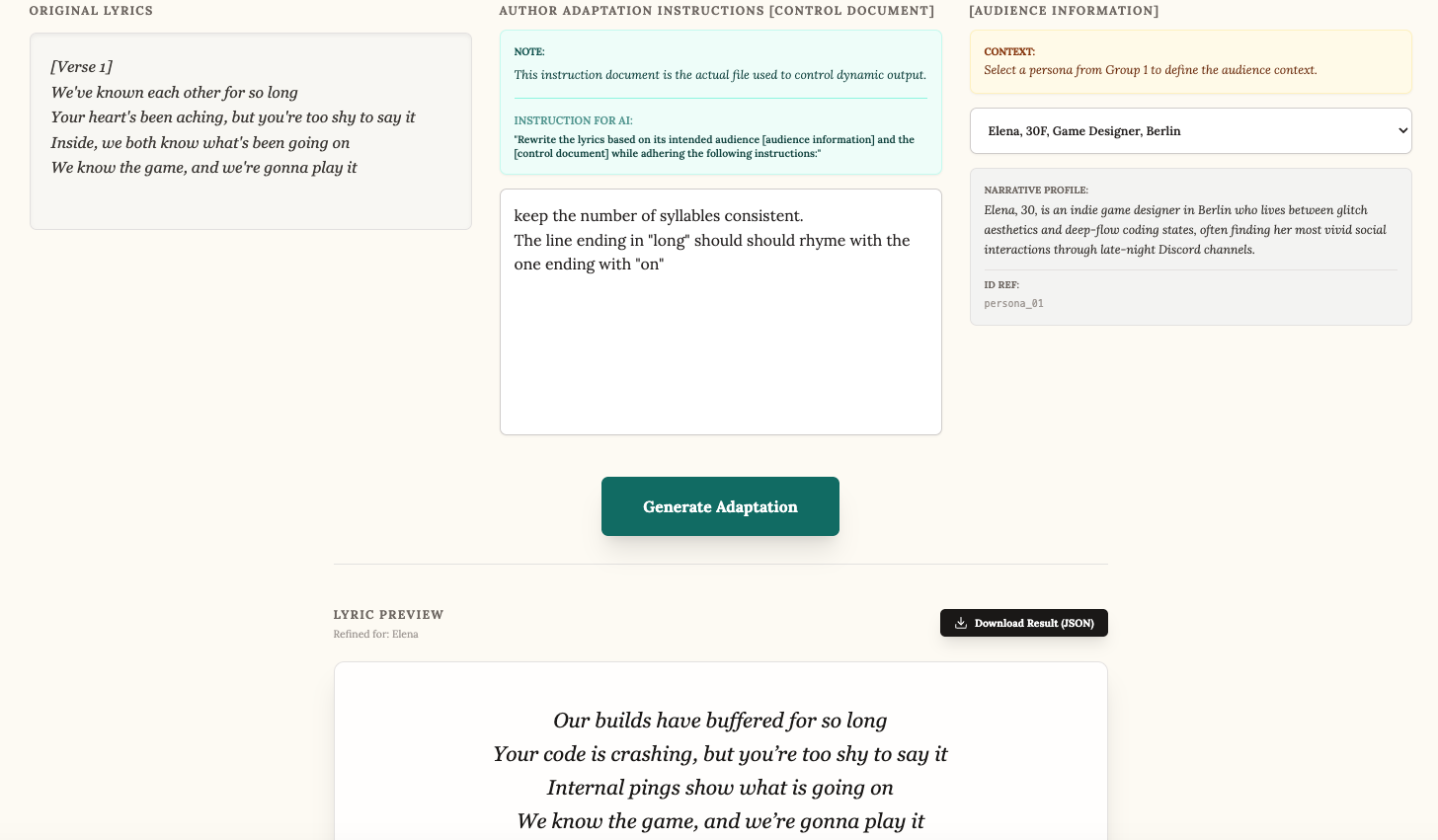}
    \caption{Interface for prompting-based workflow.}
\end{figure*}

\begin{table*}[t]
\centering
\caption{Baseline prompt template.}
\label{tab:baseline-prompt}
\begin{minipage}{0.95\textwidth}
\ttfamily\small\raggedright
Prompt = You are a songwriter, take the original lyrics and vary the lyrics while adhering to the instructions below:\par
\medskip
Rewrite the lyrics based on its intended audience \$\{audienceInfo || '[audience information]'\} and the [control document] while adhering the following instructions: \$\{instructions\}\par
\medskip
Original Lyrics: \$\{lyrics\}\par
\medskip
only output the lyrics without any additional text.
\end{minipage}
\end{table*}

\end{document}